\documentclass[12pt]{article}
\usepackage{epsfig}
\def\be{\begin{equation}}
\def\ee{\end{equation}}
\def\bea{\begin{eqnarray}}
\def\eea{\end{eqnarray}}
\usepackage{graphicx}

\catcode`\@=11
\def\lsim{\mathrel{\mathpalette\@versim<}}
\def\gsim{\mathrel{\mathpalette\@versim>}}
\def\@versim#1#2{\vcenter{\offinterlineskip
\ialign{$\m@th#1\hfil##\hfil$\crcr#2\crcr\sim\crcr } }}
\catcode`\@=12
\usepackage{axodraw}

\catcode`@=12
\begin{document}
\thispagestyle{empty}
\begin{flushright}
UCRHEP-T538\\
December 2013\
\end{flushright}
\vspace{0.6in}
\begin{center}
{\Large \bf Radiative Origin of All Quark and Lepton Masses\\ 
through Dark Matter with Flavor Symmetry\\}
\vspace{1.2in}
{\bf Ernest Ma\\}
\vspace{0.2in}
{\sl Department of Physics and Astronomy, University of California,\\
Riverside, California 92521, USA\\}
\end{center}
\vspace{1.2in}
\begin{abstract}\
The fundamental issue of the origin of mass for all quarks and leptons 
(including Majorana neutrinos) is linked to dark matter, odd under an 
exactly conserved $Z_2$ symmetry which may or may not be derivable from 
an $U(1)_D$ gauge symmetry.  The observable sector interacts with a 
proposed dark sector 
which consists of heavy neutral singlet Dirac fermions and suitably 
chosen new scalars.  Flavor symmetry is implemented in a renormalizable 
context with just the one Higgs doublet $(\phi^+,\phi^0)$ of the standard 
model in such a way that all observed fermions obtain their masses 
radiatively through dark matter.
\end{abstract}

\newpage
\baselineskip 24pt

In the standard model (SM) of particle interactions, the origin of mass is 
the electroweak Higgs doublet $(\phi^+,\phi^0)$.  With the recent 
discovery~\cite{atlas12,cms12} of the 126 GeV particle at the Large Hadron 
Collider (LHC), this is apparently confirmed as the one physical neutral 
Higgs boson, i.e. $h = \sqrt{2} Re(\phi^0)$ predicted by the SM, leaving 
the other three degrees of freedom, i.e. $(\phi^\pm,\sqrt{2} Im(\phi^0))$ 
as the longitudinal components of the observed massive electroweak vector 
gauge bosons $(W^\pm,Z^0)$.  On the other hand, the existence of dark matter 
and the observed flavor structure of quarks and leptons remain unexplained.

In this paper, it is proposed that these three fundamental issues (mass, 
dark matter, and flavor) are interconnected in the context of a theoretical 
framework for the radiative generation of all SM fermion masses.

Consider first the origin of charged-lepton masses.  Under the $SU(2)_L 
\times U(1)_Y$ gauge symmetry of the SM, there are left-handed doublets 
$L_{iL} = (\nu_i,l_i)_L$ and right-handed singlets $l_{iR}$.  The one Higgs 
doublet $\Phi = (\phi^+,\phi^0)$ of the SM connects them through the 
invariant Yukawa terms
\begin{equation}
{\cal L}_Y = f_i \bar{L}_{iL} l_{iR} \Phi + H.c. = f_i (\bar{\nu}_{iL} \phi^+ 
+ \bar{l}_{iL} \phi^0) l_{iR} + H.c.
\end{equation}
As $\phi^0$ acquires a vacuum expectation value $\langle \phi^0 \rangle = v$, 
charged leptons become
 massive with $m_i = f_i v$.

Suppose now there exists a flavor symmetry which forbids Eq.~(1).  As a 
concrete example, consider the non-Abelian discrete symmetry 
$A_4$~\cite{mr01,m02,bmv03,m04}, which is also the symmetry group of 
the tetrahedron.  It has four irreducible representations 
$\underline{1}, \underline{1}', \underline{1}'', \underline{3}$, 
with the multiplication rule
\begin{equation}
\underline{3} \times \underline{3} = \underline{1} + \underline{1}' + 
\underline{1}'' +  \underline{3} + \underline{3}.
\end{equation}
Let $L_{iL} \sim \underline{1}, \underline{1}', \underline{1}''$, 
$l_{iR} \sim \underline{3}$, $\Phi \sim \underline{1}$, then the usual SM Yukawa 
couplings are forbidden.  There are two ways now for the charged leptons 
to acquire mass. (1) Nonrenormalizable interactions of the form 
$\bar{L}_{iL} l_{jR} \Phi \chi_k$ are postulated, where the flavor structure 
is carried by the scalar singlets $\chi_k$.  For $A_4$, $\chi_k \sim 
\underline{3}$ works.  This scenario requires an ultraviolet completion, 
which may involve many new fields and parameters, and often raises more 
questions than the ones it attempts to answer. (2) Renormalizability of the 
theory is maintained by extending the scalar sector to include more 
doublets which should be observable.  For $A_4$, $\Phi_{1,2,3} \sim 
\underline{3}$ works, as proposed in the original 
papers~\cite{mr01,m02,bmv03,m04}. 

Given that the recently discovered 126 GeV particle~\cite{atlas12,cms12} 
at the LHC is very likely to be the one Higgs boson of the SM, it is 
time to consider how a renormalizable theory of flavor may be sustained 
with just the one Higgs boson of the SM.  The solution is actually very simple. 
Let all known fermion masses be generated as quantum corrections from 
a dark sector which also carries flavor.  This notion links the origin 
of SM fermion masses with flavor and dark matter which could also be 
self-interacting, an idea increasingly relevant to present astrophysical 
observations~\cite{tyz13}.

To implement this interconnected scenario, new particles have to be 
introduced.  As a simple example, three heavy {\it Dirac} singlet neutral 
fermions $N_{1,2,3}$ will be added~\cite{fm12}.  Their masses will be the 
origin of all SM 
fermion masses (including those of the Majorana neutrinos).  Consider first 
again the charged-lepton masses.  A scalar doublet $(\eta^+,\eta^0)$ 
and a charged scalar singlet $\chi^+$ are added.  These new particles 
may transform under a new $U(1)_D$ gauge symmetry, as well as $A_4$, i.e.
\begin{equation}
(\eta^+,\eta^0), ~\chi^+ \sim \underline{1}, ~~~ N_{iL} \sim \underline{3}, 
~~~ N_{iR} \sim \underline{1}, \underline{1}', \underline{1}''.
\end{equation}
The allowed Yukawa couplings are then $f'_i \bar{N}_{iR} (l_{iL} \eta^+ - 
\nu_L \eta^0)$ and $f'' \bar{l}_{iR} N_{iL} \chi^-$.  Together with the 
invariant scalar trilinear coupling $\mu (\eta^+ \phi^0 - \eta^0 \phi^+) 
\chi^-$, a radiative charged-lepton mass matrix is obtained as shown in Fig.~1, 
\begin{figure}[htb]
\vspace*{-3cm}
\hspace*{-3cm}
\includegraphics[scale=1.0]{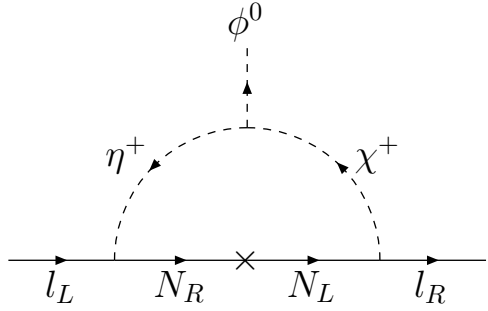}
\vspace*{-21.5cm}
\caption{One-loop generation of charged-lepton mass with $U(1)_D$ symmetry.}
\end{figure}
where the $\bar{N}_L N_R$ mass terms break $A_4$ explicitly but softly. 
Encoding of the flavor symmetry is thereby accomplished in a renormalizable 
theory, instead of the usual nonrenormalizable approach using $\bar{L}_L 
l_R \Phi \chi$.  Note that $U(1)_D$ may remain unbroken in the loop.

The nature of the soft breaking of $A_4$ is encoded in the $\bar{N}_{iL} N_{jR}$ 
mass matrix.  Let
\begin{equation}
{\cal M}_N = 
{1 \over \sqrt{3}} \pmatrix{1 & 1 & 1 \cr 1 & \omega & \omega^2 \cr 
1 & \omega^2 & \omega} \pmatrix{M_1 & 0 & 0 \cr 0 & M_2 & 0 \cr 0 & 0 & M_3} ,
\end{equation}
then a residual $Z_3$ symmetry exists~\cite{m10} which protects it 
against arbitrary corrections.  The charged-lepton mass matrix 
$\bar{l}_{iL} l_{jR}$ is then 
given by
\begin{equation}
{\cal M}_l = \pmatrix{m_e & 0 & 0 \cr 0 & m_\mu & 0 \cr 0 & 0 & m_\tau} 
{1 \over \sqrt{3}} \pmatrix{1 & 1 & 1 \cr 1 & \omega^2 & \omega 
\cr 1 & \omega & \omega^2} ,
\end{equation}
where
\begin{equation}
m_{e,\mu,\tau} = {f'' f'_{1,2,3} \sin \theta \cos \theta M_{1,2,3} \over 16 \pi^2} 
\left[ {m_1^2 \over m_1^2 - M_{1,2,3}^2} \ln {m_1^2 \over M_{1,2,3}^2} - 
{m_2^2 \over m_2^2 - M_{1,2,3}^2} \ln {m_2^2 \over M_{1,2,3}^2} \right], 
\end{equation}
with $m^2_{1,2}$ the eigenvalues and $\theta$ the mixing angle of 
the mass-squared matrix
\begin{equation}
{\cal M}^2_{\eta \chi} = \pmatrix{m_\eta^2 & \mu v \cr \mu v & m_\chi^2}.
\end{equation}
Note that ${\cal M}_l$ of Eq.~(5) is the conjugate of the result 
obtained in the original $A_4$ proposal~\cite{mr01}, using $\Phi_{1,2,3} \sim 
\underline{3}$ and $\langle \phi_1^0 \rangle = \langle \phi_2^0 \rangle 
= \langle \phi_3^0 \rangle = v/\sqrt{3}$.

At this stage, the theory is invariant under $Z_3$ and the massive 
charged leptons $(e,\mu,\tau)$, the massless neutrinos 
$(\nu_e,\nu_\mu,\nu_\tau)$, as well as the heavy dark fermion singlets 
$(N_e,N_\mu,N_\tau)$ all transform as $(1, \omega, \omega^2)$ under $Z_3$. 
The next step is to obtain the radiative generation of Majorana neutrino 
masses.  If $U(1)_D$ is replaced by $Z_2$, then the well-studied one-loop 
scotogenic model~\cite{m06} may be used.  If $U(1)_D$ is retained, then 
the recent one-loop proposal~\cite{mpr13} with two scalar doublets 
$(\eta_{1,2}^+, \eta_{1,2}^0)$ transforming oppositely under $U(1)_D$ is a 
good simple choice.  However, a two-loop realization may also be adopted, 
as shown in Fig.~2, which may preserve $U(1)_D$ as well.  Under $Z_3$, 
$\nu_{e,\mu,\tau}, N_{e,\mu,\tau}, \rho_{1,2,3} \sim 1,\omega,\omega^2$, 
$(\phi^+,\phi^0), (\eta^+,\eta^0), \chi^0 \sim 1$.  Under $U(1)_D$, $N, 
(\eta^+,\eta^0), \chi^0 \sim 1$, $\rho \sim 2$.
\begin{figure}[htb]
\vspace*{-3cm}
\hspace*{-3cm}
\includegraphics[scale=1.0]{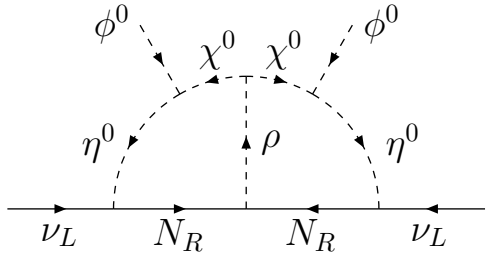}
\vspace*{-21.5cm}
\caption{Two-loop generation of Majorana neutrino mass with $U(1)_D$ symmetry.}
\end{figure}

The addition of $\chi^0$ and $\rho_1$ completes the two loops without 
breaking $U(1)_D$ or $Z_3$.  Since $(\nu_e,\nu_\mu,\nu_\tau)$ transform as 
$(1, \omega, \omega^2)$ under the unbroken residual $Z_3$ at this stage, 
this would result in a Majorana neutrino mass matrix in the basis 
$(\nu_e,\nu_\mu,\nu_\tau)$ of the form
\begin{equation}
{\cal M}_\nu = \pmatrix{A & 0 & 0 \cr 0 & 0 & B \cr 0 & B & 0}.
\end{equation}
The further addition of $\rho_{2,3}$ together with the soft breaking of $Z_3$ 
using the trilinear $\chi^0 \chi^0 \rho_{2,3}^\dagger$ couplings allows 
${\cal M}_\nu$ to become
\begin{equation}
{\cal M}_\nu = \pmatrix{A & C  & C^* \cr C & D^* & B \cr C^* & B & D},
\end{equation}
where $A$ and $B$ are real.  Note that this pattern is protected by a 
symmetry first pointed out in Ref.~\cite{gl04}, i.e. $e \to e$ and 
$\mu - \tau$ interchange with $CP$ conjugation, 
and obtained previously in Ref.~\cite{bmv03}.  As such, it is also 
guaranteed to yield maximal $\nu_\mu - \nu_\tau$ mixing ($\theta_{23} = \pi/4$) 
and maximal $CP$ violation, i.e. $\exp(-{i \delta}) = \pm i$, whereas 
$\theta_{13}$ may be nonzero and arbitrary.

The mass matrix of Eq.~(9) has six parameters: $A, B, C_R, C_I, D_R, D_I$, but 
only five are independent because the relative phase of $C$ and $D$ is 
unobservable.  Using the conventional parametrization of the neutrino 
mixing matrix, the angle $\theta_{13}$ is given by
\begin{equation}
{s_{13} \over c_{13}} = {-D_I \over \sqrt{2} C_R}, ~~~~~~ 
{s_{13} c_{13} \over c_{13}^2 - s_{13}^2} = {\sqrt{2} C_I \over A - B + D_R}.
\end{equation}
The adjustable relative phase of $C$ and $D$ is used to allow the above 
two equations to be satisfied with a single $\theta_{13}$.  The angle 
$\theta_{12}$ is then 
given by
\begin{equation}
{s_{12} c_{12} \over c_{12}^2 - s_{12}^2} = {-\sqrt{2} (c_{13}^2 - s_{13}^2) C_R 
\over c_{13} [ c_{13}^2 (A - B - D_R) + 2 s_{13}^2 D_R]}.
\end{equation}
As a result, the three mass eigenvalues are 
\begin{eqnarray}
m_1 &=& {c_{13}^2 [c_{12}^2 A - s_{12}^2 B - s_{12}^2 D_R] - s_{13}^2 
[(c_{12}^2 - s_{12}^2) B - D_R] \over (c_{13}^2 - s_{13}^2)
(c_{12}^2 - s_{12}^2)}, \\ 
m_2 &=& {c_{13}^2 [-s_{12}^2 A + c_{12}^2 B + c_{12}^2 D_R] - s_{13}^2 
[(c_{12}^2 - s_{12}^2) B + D_R] \over (c_{13}^2 - s_{13}^2)
(c_{12}^2 - s_{12}^2)}, \\ 
m_3 &=& {s_{13}^2 A - c_{13}^2 B + c_{13}^2 D_R \over c_{13}^2 - s_{13}^2}.
\end{eqnarray}
Since $s_{13}^2 \simeq 0.025$ is small, these expressions become
\begin{eqnarray}
m_2 + m_1 &\simeq& A + B + D_R + s_{13}^2 (A - B + D_R), \\ 
(c_{12}^2 - s_{12}^2) (m_2 - m_1) &\simeq& -A + B + D_R - s_{13}^2 (A - B + D_R), \\ 
m_3 &\simeq& -B + D_R + s_{13}^2 (A - B + D_R).
\end{eqnarray}
It is clear that a realistic neutrino mass spectrum with $m_2^2 - m_1^2 << 
|m_3^2 - (m_2^2 + m_1^2)/2|$ may be obtained with either 
$|m_1| < |m_2| < |m_3|$ (normal ordering) or $|m_3| < |m_1| < |m_2|$ 
(inverted ordering).

As for quarks, color triplet scalars are added, i.e. the electroweak doublet 
$(\xi^{2/3}, \xi^{-1/3})$ and singlets $\zeta^{2/3}$, $\zeta^{-1/3}$, all 
transforming as $-1$ under $U(1)_D$.  The analogs of Fig.~1 are easily 
obtained for ${\cal M}_{u,d}$ \underline{using again just $N_{1,2,3}$}, 
as shown in Figs.~3 and 4.  
\begin{figure}[htb]
\vspace*{-3cm}
\hspace*{-3cm}
\includegraphics[scale=1.0]{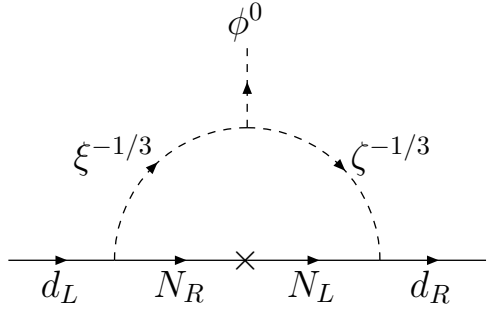}
\vspace*{-21.5cm}
\caption{One-loop generation of $d$ quark mass with $U(1)_D$ symmetry.}
\end{figure}
\begin{figure}[htb]
\vspace*{-3cm}
\hspace*{-3cm}
\includegraphics[scale=1.0]{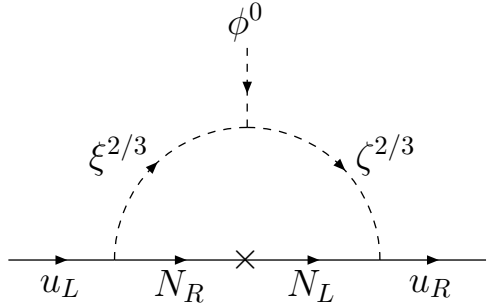}
\vspace*{-21.5cm}
\caption{One-loop generation of $u$ quark mass with $U(1)_D$ symmetry.}
\end{figure}
Under the symmetry $A_4$ again as an example,
\begin{eqnarray}
&& Q_{iL} = (u,d)_{iL} \sim \underline{1}, \underline{1}', \underline{1}'', 
~~~ u_{iR},~d_{iR} \sim \underline{3}, \\ 
&& (\xi^{2/3}, \xi^{-1/3}), ~\zeta^{2/3}, ~\zeta^{-1/3} \sim \underline{1},
\end{eqnarray}
in complete analogy to the charged-lepton sector, resulting also in 
arbitrary quark masses with a residual symmetry $Z_3$ under which  
$(u,c,t)$ and $(d,s,b)$ transform as $(1,\omega,\omega^2)$.
(Note that to get a realistic $m_t$ using Eq.~(6) with Yukawa couplings of 
order unity, $M_3$ and $m_{1,2}$ should all be of order 10 TeV, with no 
cancellation between the $m_{1,2}$ terms.  A variant of this scheme is to use 
a flavor symmetry such that only $t$ couples to $\Phi$ at tree level, 
which may be possible with $\Delta(27)$ for example, because it has 9 
inequivalent one-dimensional representations.)  The quark matrices are thus 
both diagonal, so there is no mixing in the quark sector in this symmetry 
limit.  In other words, there is now a theoretical understanding of 
why the actual mixing angles are small.  They are the result of breaking 
this $Z_3$ symmetry in the soft terms of the Dirac mass matrix of the 
singlet $N$'s.

The $\bar{q}_{iL} q_{jR}$ mass matrix is of the form
\begin{equation}
{\cal M}_q = \pmatrix{f_1 & 0 & 0 \cr 0 & f_2 & 0 \cr 0 & 0 & f_3} 
U_M^L \pmatrix{M_1 & 0 & 0 \cr 0 & M_2 & 0 \cr 0 & 0 & M_3} (U_M^R)^\dagger.
\end{equation}
If the unitary matrices $U_M^{L,R}$ are the identity, then $Z_3$ is not broken 
and the quark mixing matrix $V_{CKM}$ is also the identity.  Let $U_M^L$ be 
approximately given by
\begin{equation}
U_M^L \simeq \pmatrix{1 & -\epsilon_{12} & -\epsilon_{13} \cr \epsilon_{12}^* 
& 1 & -\epsilon_{23} \cr \epsilon_{13}^* & \epsilon_{23}^* & 1}
\end{equation}
then 
\begin{equation}
{\cal M}_q {\cal M}_q^\dagger \simeq \pmatrix{f_1^2 M_1^2 & f_1 f_2 \epsilon_{12} 
(M_1^2 - M_2^2) & f_1 f_3 \epsilon_{13} (M_1^2 - M_3^2) \cr f_1 f_2  \epsilon_{12}^* 
(M_1^2 - M_2^2) & f_2^2 M_2^2 & f_2 f_3 \epsilon_{23} (M_2^2 - M_3^2) \cr 
f_1 f_3 \epsilon_{13}^* (M_1^2 - M_3^2) & f_2 f_3 \epsilon_{23}^* (M_2^2 - M_3^2) 
& f_3^2 M_3^2}.
\end{equation}
Let $m_d \simeq f_1^d M_1$, $m_s \simeq f_2^d M_2$, $m_b \simeq f_3^d M_3$, 
$m_u \simeq f_1^u M_1$, $m_c \simeq f_2^u M_2$, $m_t \simeq f_3^u M_3$, 
then $V_{CKM}$ is approximately 
given by
\begin{eqnarray}
&& V_{ud} \simeq V_{cs} \simeq V_{tb} \simeq 1, ~~~ 
V_{us} \simeq \left( {m_d \over m_s} - {m_u \over m_c} \right) \epsilon_{12} 
\left( {M_2^2 - M_1^2 \over M_2 M_1} \right), \\
&& V_{ub} \simeq \left( {m_d \over m_b} - {m_u \over m_t} \right) \epsilon_{13} 
\left( {M_3^2 - M_1^2 \over M_3 M_1} \right), ~~~ 
V_{cb} \simeq \left( {m_s \over m_b} - {m_c \over m_t} \right) \epsilon_{23} 
\left( {M_3^2 - M_2^2 \over M_3 M_2} \right).
\end{eqnarray}
There are many realistic solutions of the above.  The simplest is to 
set $f_1^d = f_2^d = f_3^d$, in which case $V_{CKM} \simeq (U_M^L)^\dagger$.  
In other words, the soft breaking of $Z_3$ which generates $U_M^L$ is directly 
linked to the observed $V_{CKM}$ .

In this grand scheme, all SM fermions owe their masses not just to the one 
and only one Higgs boson, but also to the invariant masses of the three neutral 
singlet fermions $N_{1,2,3}$.  Flavor structure is carried by the $\bar{N}_{iL} 
N_{jR}$ mass matrix and transmitted to the quarks and leptons through new 
observable scalars.  These scalars as well as $N_{1,2,3}$ may have their own 
$U(1)_D$ gauge interactions, and the lightest $N$ is a dark-matter 
candidate.  The $U(1)_D$ gauge symmetry may be broken to or replaced 
by an exact residual $Z_2$ symmetry which maintains the stability of dark 
matter.    A specific $A_4$ 
model of flavor has been presented which shows how a predictive neutrino 
mass matrix may be generated in two loops, and how a realistic $V_{CKM}$ 
matrix is obtained using again only $N_{1,2,3}$.  There are no flavor-changing 
neutral-current (FCNC) processes at tree level.  They may appear in one 
loop through the dark sector, but they are suppressed by the flavor symmetry.  
The interconnection between mass, flavor, and dark matter is the key.

Whereas the SM fermions are known to transform as $\underline{5}^*$ and 
$\underline{10}$ of $SU(5)$, the new dark-sector scalars do so as well:
\begin{equation}
(\eta^+,\eta^0), \zeta^{-1/3} \sim \underline{5}, ~~~~~ 
(\xi^{2/3},\xi^{-1/3}), (\zeta^{2/3})^*, \chi^+ \sim \underline{10}.
\end{equation}
In a supersymmetric context, these would be squarks and sleptons. 
Analogous diagrams to Figs.~1, 3, and 4 have been previously 
discussed~\cite{m89} in this regard.  Instead of $R$ parity, they are 
distinguished here by dark $Z_2$.  Note that $U(1)_D$ is not compatible 
with this interpretation.  The color triplet $\zeta^{-1/3}$ has also been 
previously considered~\cite{m08}.  As for the singlets $N_{1,2,3}, \rho_{1,2,3}$, 
and $\chi^0$, they are also singlets under $SU(5)$, although they may also 
have an $SU(6)$ origin~\cite{m13}.

The renormalization-group (RG) equations for the evolution of the SM gauge 
couplings are equally affected by the new particles because they form 
complete $SU(5)$ multiplets, so there is no gauge-coupling unification 
as in the SM, but the simple addition of a few new particles will do 
the trick~\cite{m05} if desired.

The new scalar particles of Eq.~(25) mimic those of supersymmetry, so they 
may be produced at the LHC.  They also have similar signatures because 
the lightest $N$ behaves as the LSP (lightest supersymmetric particle) of 
the MSSM (minimal supersymmetric standard model).  However, there is no 
gluino in this theory, so details of the quark-squark interactions will 
be different. 

The lightest $N$ is the dark-matter candidate of this proposal.  Since it 
is a Dirac fermion, its annihilation cross section to quark and lepton 
pairs through the exchange of scalar quarks and leptons are unsuppressed, 
unlike the case of the MSSM with the lightest Majorana neutralino as the LSP.  
Its couplings are also not constrained by the MSSM.  Hence it has a much 
larger parameter space to be a viable dark-matter candidate and has greater 
discovery potential at the LHC.  The phenomenology of such a Dirac fermion 
dark-matter candidate has been discussed in Ref.~\cite{fm12,mpr13}.  Its 
relic density has been shown to be compatible with what is observed.  Here 
there are more annihilation channels, but an overall acceptable 
parameter space is clearly available.

In summary, a unifying proposal has been made. (1) In addition to the SM 
particles, there exists a dark sector, odd under $Z_2$ which may be derived 
from an $U(1)_D$ gauge symmetry.  The particles of this dark sector 
consist of three neutral singlet Dirac fermions $N_{1,2,3}$ and the scalars 
of Eq.~(25) which are complete $SU(5)$ multiplets.  The lightest $N$ is 
a possible dark-matter candidate.  (2) A non-Abelian discrete symmetry 
such as $A_4$ exists, under which $N_{1,2,3}$ as well as the quarks and 
leptons of the SM transform nontrivially.  (3) As a result of this flavor 
symmetry, the one and only one Higgs doublet $\Phi$ of the SM is forbidden 
to couple to $\bar{q}_L q_R$ and $\bar{l}_L l_R$.  The nonzero vacuum 
expectation value of $\phi^0$ generates $W$ and $Z$ masses but not fermion 
masses.  (4) The soft breaking of $A_4$ to $Z_3$ in the $3 \times 3$ Dirac 
mass matrix of $N_{1,2,3}$ allows $\Phi$ to couple to $\bar{q}_L q_R$ and 
$\bar{l}_L l_R$ in one loop.  Thus all quarks and leptons owe their masses 
to dark matter in conjunction with $\Phi$.  (5) The residual $Z_3$ symmetry 
maintains diagonal mass matrices for $u$ and $d$ quarks as well as charged 
leptons. This is an explanation of why the quark mixing matrix $V_{CKM}$ is 
nearly diagonal.  (6) Further soft breaking of $Z_3$ allows a realistic 
$V_{CKM}$. A two-loop Majorana neutrino mass matrix is also obtained with the 
addition of scalar singlets $\chi$ and $\rho_{1,2,3}$.  (7) The resulting 
neutrino mass matrix may be implemented with a generalized $CP$ transformation 
under $\mu-\tau$ exchange to obtain maximal $CP$ violation together with 
$\theta_{23} = \pi/4$ while allowing nonzero $\theta_{13}$.  (8) The 
predicted scalars of Eq.~(25) which connect the quarks and leptons to 
their common dark-matter antecedents, i.e. $N_{1,2,3}$, are possibly 
observable at the LHC.  They may also change significantly the SM couplings 
of $\Phi$~\cite{fm13}.

\medskip

This work is supported in part 
by the U.~S.~Department of Energy under Grant No.~DE-SC0008541.

\baselineskip 18pt

\bibliographystyle{unsrt}

\end{document}